\documentclass[pra,aps,superscriptaddress,showpacs]{revtex4}
\usepackage{graphicx,graphics,epsfig}
\usepackage{dcolumn}
\usepackage{bm}
\usepackage{amsmath}
\usepackage{verbatim}
\usepackage{color}
\usepackage[colorlinks=false]{hyperref} 
\usepackage{subfigure}
\usepackage{times,natbib}
\usepackage{amsmath,amsfonts,amssymb,graphics,graphicx,epsfig,color,times,natbib}
\usepackage{tikz}
\usepackage{pgfplots}
\usepackage{verbatim}
\usepackage{fancyhdr} 
\begin{document}
\title{More about the doubling degeneracy operators associated with Majorana fermions\\ and Yang-Baxter equation}

\author{Li-Wei Yu}
\email{NKyulw@gmail.com}
\affiliation{Theoretical Physics Division, Chern Institute of Mathematics, Nankai University,
Tianjin 300071, China}

\author{Mo-Lin Ge}
\affiliation{Theoretical Physics Division, Chern Institute of Mathematics, Nankai University,
Tianjin 300071, China}



\begin{abstract}

\textbf{A new realization of doubling degeneracy based on emergent Majorana operator $\Gamma$ presented by Lee-Wilczek has been made. The Hamiltonian  can be obtained through the new type of solution of Yang-Baxter equation, i.e. $\breve{R}(\theta)$-matrix. For 2-body interaction, $\breve{R}(\theta)$ gives the ``superconducting''  chain that is the same as 1D Kitaev chain model. The 3-body Hamiltonian commuting with $\Gamma$ is derived by 3-body $\breve{R}_{123}$-matrix, we thus show that the essence of the doubling degeneracy is due to $[\breve{R}(\theta), \Gamma]=0$. We also show that the extended $\Gamma'$-operator is an invariant of braid group $B_N$ for odd $N$. Moreover, with the extended $\Gamma'$-operator, we construct the high dimensional matrix representation of solution to Yang-Baxter equation and find its application in constructing $2N$-qubit Greenberger-Horne-Zeilinger state for odd $N$. }

\end{abstract}

\pacs{71.10.Pm,
74.20.Mn,
02.10.De,
03.67.Mn,
}

\maketitle


The Majorana mode \cite{kitaev2001unpaired,ivanov2001non,wilczek2009majorana,leijnse2012introduction} has attracted increasing attention in physics due to its potential applications in topological quantum information processing \cite{kitaev2003fault,kitaev2006anyons,alicea2011non}. Specifically, the degenerate ground state in Majorana mode serves as topologically protected states which can be used for topological quantum memory.

\medskip

 In the Ref. \cite{lee2013algebra}, Lee and Wilczek presented a new operator $\Gamma$ that provided the doubling degeneracy for the Hamiltonian formed by Majorana fermions to overcome the  conceptional incompletion of the algebraic set for the Majorana model. Following the Ref. \cite{lee2013algebra}, the Majorana operators $\gamma_i$'s satisfy Clifford algebraic relations:
\begin{equation}\label{CliffordAlgebra}
\left\{ \gamma_{i},\gamma_{j}\right\} =2\delta_{ij},
\end{equation}
and the Hamiltonian takes the form
\begin{equation}
H_{\textrm{int}}=-\textrm{i}\left(\alpha \gamma_{1}\gamma_{2}+\beta \gamma_{2}\gamma_{3}+\kappa \gamma_{3}\gamma_{1}\right). \label{Eq.2}
\end{equation}
The algebra in equation (\ref{CliffordAlgebra}) is conceptually incomplete. Besides the parity, the
nonlinear operator $\Gamma$ is introduced \cite{lee2013algebra}

\begin{equation}
\Gamma=-\textrm{i}\gamma_{1}\gamma_{2}\gamma_{3}
\end{equation}
to form the set
\begin{eqnarray}
&\Gamma^{2}=1,   P^{2}=1, \left[\Gamma,H_{\textrm{int}}\right]=0, \left[P,H_{\textrm{int}}\right]=0, \\
&\left[\Gamma,\gamma_{j}\right]=0, \left\{ P, \gamma_{j}\right\} =0 , \,\left\{ \Gamma,P\right\} =0, \nonumber
\end{eqnarray}
where  $P$ implements the electron number parity, and $P^2=1$. The emergent Majorana operator $\Gamma$ and  parity operator $P$ lead to the doubling degeneracy at any energy level, not only for the ground state.

On the other hand, based on the obtained new type of solution $\breve{R}_i(\theta)$ of Yang-Baxter equation (YBE), which is related to Majorana operators, the corresponding Hamiltonian can be found by following the standard way \cite{baxter1972partition}, i.e. the Hamiltonian $H\sim\frac{\partial\breve{R}_i(\theta)}{\partial\theta}|_{\theta=0}$. We find that the Hamiltonian derived from $\breve{R}_i(\theta)$ is 1D Kitaev model \cite{kitaev2001unpaired}. Moreover, because 1+1D 3-body S-matrix can be decomposed into three 2-body S-matrices based on YBE,  we construct the 3-body Hamiltonian from 3-body S-matrix and find its doubling degeneracy. Hence, the advantage of parametrizing the braiding operator $B_i$ to $\breve{R}_i(\theta)$ is that the desired Hamiltonian associated with Majorana operators can be derived from $\breve{R}_i(\theta)$.

Now let us first give a brief introduction to the Majorana representation of  braiding operator as well as the solution of Yang-Baxter equation.

The non-Abelian statistics \cite{Anyonstatistics} of Majorana fermion (MF) has been proposed in both 1D quantum wires network \cite{alicea2011non} and 2D $p+\textrm{i}p$ superconductor \cite{ivanov2001non}.

For $2N$ Majorana fermions, the braiding operators of Majorana fermions form braid group $B_{2N}$ generated by elementary interchanges $B_{i}=U_{i,i+1}=\exp(\tfrac{\pi}{4}\gamma_i\gamma_{i+1})$ of neighbouring particles ($i=1,2\cdots 2N-1$) with the following braid relations:
\begin{eqnarray}
  B_i B_{i+1}B_i=B_{i+1}B_iB_{i+1}, \\
  B_i B_j=B_j B_i, \quad |i-j|>1.
\end{eqnarray}

The Yang-Baxter equation (YBE) \cite{yang1967some,yang1968matrix,baxter1972partition} is a natural generalization of braiding relation with the parametrized form:
\begin{equation}\label{YBEMultiply}
\breve{R}_{i}(x) \breve{R}_{i+1}(xy) \breve{R}_{i}(y)=\breve{R}_{i+1}(y) \breve{R}_{i}(xy) \breve{R}_{i+1}(x),
\end{equation}
where $x$, $y$ stand for spectral parameters,
\begin{equation}
\breve{R}_{i}=\tfrac{1}{\sqrt{1+x^2}}(B_{i}+xB_{i}^{-1}).
\end{equation}

The solutions of equation (\ref{YBEMultiply}) was intensively studied by Yang, Baxter, Faddeev and other authors     \cite{yang1967some,yang1968matrix,baxter2007exactly,batchelor2007bethe,yang1991braid,takhtadzhan1979quantum,faddeev1981soviet,kulish1981lecture,kulish1982solutions,korepin1997quantum} in dealing with many body problems, statistical models, low-dimensional quantum field theory, spin chain models and so on. We call this type of solutions type-I.

Based on Ref. \cite{hietarinta1993solving} there appears a new type of solutions called type-II \cite{kauffman2002quantum,chen2007braiding,hu2008optical,ge2012yang}. By introducing a new variable  $\theta$ as $\cos\theta=\tfrac{1+x}{\sqrt{2(1+x^2)}}$ and
$\sin\theta=\tfrac{1-x}{\sqrt{2(1+x^2)}}$, we have
\begin{equation}\label{MajoranaSolution}
\breve{R}_i(\theta)=e^{\theta\gamma_i\gamma_{i+1}}=\cos\theta+\sin\theta\gamma_i\gamma_{i+1},
\end{equation}
then the YBE reads \cite{jimbobook}:
\begin{equation}\label{YBEtheta}
\breve{R}_{i}(\theta_1) \breve{R}_{i+1}(\theta_2) \breve{R}_{i}(\theta_3)=\breve{R}_{i+1}(\theta_3) \breve{R}_{i}(\theta_2) \breve{R}_{i+1}(\theta_1),
\end{equation}
with the constraint for three parameters $\theta_1$, $\theta_2$ and $\theta_3$ :
\begin{equation}\label{YBEangularrelation}
\tan\theta_2=\frac{\tan\theta_1+\tan\theta_3}{1+\tan\theta_1\tan\theta_3},
\end{equation}
i.e. the Lorentzian additivity by  $\tan\theta=\frac{1}{c}u$. It is well known that the physical meaning of $\theta$ is to describe entangling degree, which is $|\sin2\theta|$ for 2-qubit \cite{chen2007braiding}.  The type-II solution of YBE $\breve{R}_i(\theta)$ means the operation between two Majorana fermions, $\gamma_i$ and $\gamma_{i+1}$. Because $\gamma_i$'s satisfy Clifford algebraic relations:
\begin{equation}
\{\gamma_i, \gamma_j\}=2\delta_{ij}.
\end{equation}
Then the solution $\breve{R}_i(\theta)=e^{\theta\gamma_i\gamma_{i+1}}$  transforms the Majorana fermions $\gamma_i$ and $\gamma_{i+1}$ in the following way:
\begin{eqnarray}
\breve{R}_i(\theta)\gamma_i\breve{R}_i^{\dagger}(\theta)&=&\cos2\theta\gamma_i-\sin2\theta\gamma_{i+1},\\
\breve{R}_i(\theta)\gamma_{i+1}\breve{R}_i^{\dagger}(\theta)&=&\sin2\theta\gamma_{i}+\cos2\theta\gamma_{i+1}.
\end{eqnarray}

Since the solution of Yang-Baxter equation can be expressed in Majorana form,  the following problems arise: (i) How to understand the $\Gamma$-operator intuitively on the basis of the concrete MF model generated by YBE; (ii) How to obtain the 3-body Hamiltonian, which possesses the doubling degeneracy, from YBE; (iii)  What is the relationship between $\Gamma$-operator (as well as extended $\Gamma'$) and the solution $\breve{R}_i(\theta)$ of YBE.

\medskip

In this paper, we show that the emergent Majorana operator $\Gamma$ is a new symmetry of $\breve{R}(\theta)$ as well as Yang-Baxter equation. Due to the symmetry, the 3-body Hamiltonian derived from YBE holds Majorana doubling.  We also present a new realization of doubling degeneracy for Majorana mode. Moreover, we discuss the topological phase in the ``superconducting'' chain. The generation of Greenberger-Horne-Zeilinger (GHZ) state  via the approach of YBE is also discussed.
\section*{RESULTS}

\noindent \textbf{Topological phase in the derived ``superconducting'' chain.}
The topological phase transition in the derived ``superconducting'' chain based on YBE is discussed. We find that our chain model is exactly the same as 1D Kitaev model. Let us first give a brief introduction to 1D Kitaev model.

1D Kitaev's toy model is one of the simplest but the most representative model for Majorana mode \cite{kitaev2001unpaired,leijnse2012introduction}. The model is a quantum wire with N sites lying on the surface of three dimensional $p$-wave superconductor, and each site is either empty or occupied by an electron with a fixed spin direction. Then the Hamiltonian is expressed as the following form:
\begin{eqnarray}\label{ToyModel}
\hat{H}_k = \sum_j^N &&[-\mu\left(a_j^{\dag}a_{j}-\tfrac{1}{2}\right)-\omega\left(a_j^{\dag}a_{j+1}+a_{j+1}^{\dag}a_{j}\right)\nonumber\\
&&+\Delta a_ja_{j+1}+\Delta^{\ast}a_{j+1}^{\dag}a_{j}^{\dag}].
\end{eqnarray}
Here $a_j^{\dag}$, $a_j$ represent spinless ordinary fermion, $\omega$ is hopping amplitude, $\mu$ is chemical potential, and $\Delta=|\Delta|e^{-i\varphi}$ is induced superconducting gap. Define Majorana fermion operators:
\begin{eqnarray}\label{MFs}
 \gamma_{2j-1}&=&e^{i\tfrac{\varphi}{2}}a_j^{\dag}+e^{-i\tfrac{\varphi}{2}}a_j,\label{MFs}\\
 \gamma_{2j}&=&i e^{i\tfrac{\varphi}{2}}a_j^{\dag}-i e^{-i\tfrac{\varphi}{2}}a_j,\label{MFs2}
 \end{eqnarray}
which satisfy the relations:
\begin{equation}
\gamma_{m}^{\dag}=\gamma_{m}, \quad \{\gamma_{l}, \gamma_{m}\}=2\delta_{lm}, \quad l,m=1,\ldots 2N.
\end{equation}
Then the Hamiltonian is transformed into the Majorana form:
\begin{eqnarray}
\hat{H}_k=\tfrac{\textrm{i}}{2}\sum_j&& [-\mu \gamma_{2j-1}\gamma_{2j}+\left(\omega+|\Delta|\right)\gamma_{2j}\gamma_{2j+1}\nonumber\\
&&+\left(-\omega+|\Delta|\right)\gamma_{2j-1}\gamma_{2j+2}].
\end{eqnarray}

An interesting case is $\mu=0$, $\omega=|\Delta|$. In this case, the Hamiltonian turns into Majorana mode corresponding to topological phase:
\begin{equation}
\hat{H}_k=\textrm{i}\omega\sum_{j} \gamma_{2j}\gamma_{2j+1}.
\end{equation}
The above Hamiltonian has two degenerate ground states, $|0\rangle$ and $|1\rangle=d^{\dag}|0\rangle$. Here $d^{\dag}=e^{-\textrm{i}\varphi/2}(\gamma_1-\textrm{i}\gamma_{2N})/2$ is a non-local ordinary fermion. The degenerate states can be used for topological quantum memory qubits that are immune to local errors.  

\bigskip

Now let us construct the ``superconducting'' chain based on the solution $\breve{R}_i(\theta)$ of YBE. We imagine that a unitary evolution is governed by $\breve{R}_i(\theta)$. If only $\theta$ in unitary operator $\breve{R}_i(\theta)$ is time-dependent, we can express a state $|\psi(t)\rangle$ as $|\psi(t)\rangle=\breve{R}_i|\psi(0)\rangle$. Taking the Schr\"{o}dinger equation $i\hbar\tfrac{\partial}{\partial t}|\psi(t)\rangle=\hat{H}(t)|\psi(t)\rangle$ into account, one obtains:
\begin{equation}
i\hbar\tfrac{\partial}{\partial t}[\breve{R}_i|\psi(0)\rangle]=\hat{H}(t)\breve{R}_i|\psi(0)\rangle.
\end{equation}
Then the Hamiltonian $\hat{H}_i(t)$ related to the unitary operator $\breve{R}_i(\theta)$ is obtained:
\begin{equation}\label{SchrodingerEquation}
\hat{H}_i(t)=\textrm{i}\hbar\tfrac{\partial\breve{R}_i}{\partial t}\breve{R}_{i}^{-1}.
\end{equation}
Substituting $\breve{R}_i(\theta)=\exp(\theta\gamma_i\gamma_{i+1})$ into equation (\ref{SchrodingerEquation}), we have:
\begin{equation}\label{2MFHamiltonian}
\hat{H}_i(t)=\textrm{i}\hbar\dot{\theta}\gamma_{i}\gamma_{i+1}.
\end{equation}
This Hamiltonian describes the interaction between $i$-th and $(i+1)$-th sites with the parameter $\dot{\theta}$. Indeed, when $\theta=n \times \tfrac{\pi}{4}$, the unitary evolution corresponds to the braiding progress of two nearest Majorana fermion sites in the system, here n is an integer and signifies  the times of braiding operation.

If we only consider the nearest-neighbour interactions between MFs and extend equation (\ref{2MFHamiltonian}) to an inhomogeneous chain with 2N sites, the derived ``superconducting''  chain model is expressed as:
\begin{equation}\label{YBEKitaev}
\hat{H}=\textrm{i}\hbar\sum_{k=1}^{N}(\dot{\theta}_1\gamma_{2k-1}\gamma_{2k}+\dot{\theta}_2\gamma_{2k}\gamma_{2k+1}),
\end{equation}
with $\dot{\theta}_1$  and $\dot{\theta}_2$ describing odd-even and even-odd pairs, respectively.

\begin{figure}[!ht]
\centering
\vspace{30mm}
\centerline{\includegraphics[scale=1]{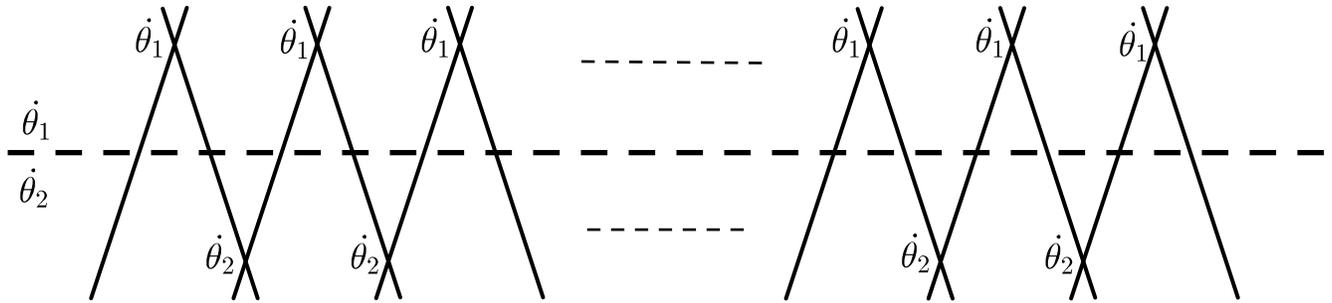}}
 \caption{{\bf The nearest neighbouring interactions of 2N Majorana sites described by the ``superconducting'' chain.} Each solid line represents a Majorana site, and the crossing means the interaction. The dashed line divides the interactions into two parts that are described by $\dot{\theta}_1$ and $\dot{\theta}_2$ respectively. When $\dot{\theta}_1=0,\dot{\theta}_2\neq0$, the first line and the last line are free, and the Hamiltonian corresponds to topological phase.}\label{Figure1}
\end{figure}

Now we give a brief discussion about the above chain model in two cases (see Fig.\ref{Figure1}):

\begin{enumerate}
\item $\dot{\theta}_1>0$, $\dot{\theta}_2=0.$

In this case, the Hamiltonian is:
\begin{equation}\label{YBEtrivil}
\hat{H}_1=\textrm{i}\hbar\sum_{k}^{N}\dot{\theta}_1\gamma_{2k-1}\gamma_{2k}.
\end{equation}
As defined in equation (\ref{MFs}) and (\ref{MFs2}), the Majorana operators $\gamma_{2k-1}$ and $\gamma_{2k}$ come from the same ordinary fermion site k, $\textrm{i}\gamma_{2k-1}\gamma_{2k}=2a_{k}^{\dag}a_{k}-1$ ($a_{k}^{\dag}$ and $a_{k}$ are spinless ordinary fermion operators). $\hat{H}_1$ simply means the total occupancy of ordinary fermions in the chain and has U(1) symmetry, $a_j\rightarrow e^{i\phi}a_j $.  Specifically, when $\theta_1(t)=\tfrac{\pi}{4}$, the unitary evolution $e^{\theta_{1}\gamma_{2k-1}\gamma_{2k}}$ corresponds to the braiding operation of two Majorana sites from the same k-th ordinary fermion site.  The ground state represents the ordinary fermion occupation number 0. In comparison to 1D Kitaev model, this Hamiltonian corresponds to the trivial case of Kitaev's. In Fig.\ref{Figure1}, this Hamiltonian is described by the intersecting lines above the dashed line, where the intersecting lines correspond to interactions. The unitary evolution of the system $e^{-i{\int\hat{H}_1dt}}$ stands for the exchange process of odd-even Majorana sites.

\item $\dot{\theta}_1=0$, $\dot{\theta}_2>0.$

In this case, the Hamiltonian is:
\begin{equation}\label{YBEtopo}
\hat{H}_2=\textrm{i}\hbar\sum_{k}^{N}\dot{\theta}_2\gamma_{2k}\gamma_{2k+1}.
\end{equation}
This Hamiltonian corresponds to the topological phase of 1D Kitaev model and has $\mathbb{Z}_2$ symmetry, $a_j\rightarrow -a_j$. Here the operators $\gamma_1$ and $\gamma_{2N}$ are absent in $\hat{H}_2$, which is illustrated by the crossing under the dashed line in Fig.\ref{Figure1}. The Hamiltonian has two degenerate ground state, $|0\rangle$ and $|1\rangle=d^{\dag}|0\rangle$, $d^{\dag}=e^{-i\varphi/2}(\gamma_1-i\gamma_{2N})/2$. This mode is the so-called Majorana mode in 1D Kitaev chain model. When $\theta_2(t)=\tfrac{\pi}{4}$, the unitary evolution $e^{\theta_{2}\gamma_{2k}\gamma_{2k+1}}$ corresponds to the braiding operation of two Majorana sites $\gamma_{2k}$ and $\gamma_{2k+1}$ from $k$-th and $(k+1)$-th ordinary fermion sites, respectively.

\medskip
\end{enumerate}

Thus we conclude that our Hamiltonian derived from $\breve{R}_{i}(\theta(t))$ corresponding to the braiding of nearest Majorana fermion sites is exactly the same as the 1D wire proposed by Kitaev, and $\dot{\theta}_1=\dot{\theta}_2$ corresponds to the phase transition point in the ``superconducting'' chain. By choosing different time-dependent parameter $\theta_1$ and $\theta_2$, we find that the Hamiltonian $\hat{H}$ corresponds to different phases.

\bigskip

\noindent \textbf{New realization of Majorana Doubling based on $\Gamma$-operator.}
The important progress had been made to establish the complete algebra for the Majorana doubling by introducing the emergent Majorana operator $\Gamma$  \cite{lee2013algebra}:

\begin{equation}
\Gamma=-\textrm{i}\gamma_{1}\gamma_{2}\gamma_{3}.
\end{equation}
 In Ref. \cite{lee2013algebra}, the concreted realization of the operators was presented in terms of Pauli matrices. On the other hand, as pointed out
in Ref. \cite{kauffman2004braiding}, there is the transformation between the natural basis and Bell basis for

\begin{eqnarray}
|\Phi_{0}\rangle&=&(|\downarrow\downarrow\rangle,|\uparrow\downarrow\rangle,|\downarrow\uparrow\rangle,|\uparrow\uparrow\rangle)^{T},\\
|\Psi\rangle&=&(|\Psi_{+}\rangle,|\Phi_{+}\rangle,|\Phi_{-}\rangle,|\Psi_{-}\rangle)^{T},
\end{eqnarray}
where
\begin{eqnarray}
\bigl|\Psi_{+}\bigr\rangle=\frac{1}{\sqrt{2}}\bigl(\bigl|\bigl\uparrow\bigr\uparrow\bigr\rangle+\bigl|\bigl\downarrow\bigr\downarrow\bigr\rangle\bigr),\quad & \bigl|\Phi_{+}\bigr\rangle=\frac{1}{\sqrt{2}}\bigl(\bigl|\bigr\uparrow\bigr\downarrow\bigr\rangle+\bigl|\bigl\uparrow\bigr\downarrow\bigr\rangle\bigr),\\
\bigl|\Psi_{-}\bigr\rangle=\frac{1}{\sqrt{2}}\bigl(\bigl|\bigl\downarrow\bigr\uparrow\bigr\rangle-\bigl|\bigl\uparrow\bigl\downarrow\bigr\rangle\bigr), \quad & \bigl|\Phi_{-}\bigr\rangle=\frac{1}{\sqrt{2}}\bigl(\bigl|\bigl\uparrow\bigr\uparrow\bigr\rangle-\bigl|\bigl\downarrow\bigr\downarrow\bigr\rangle\bigr)
\end{eqnarray}
through the matrix $B_{II}$:
\begin{equation}
\bigl|\Psi\bigr\rangle=B_{II}\bigl|\Phi_{0}\bigr\rangle,
\end{equation}
where
\begin{equation}
B_{II}=\frac{1}{\sqrt{2}}\left[\begin{array}{cccc}
1 & 0 & 0 & 1\\
0 & 1 & 1 & 0\\
0 & -1 & 1 & 0\\
-1 & 0 & 0 & 1
\end{array}\right]=\frac{1}{\sqrt{2}}\bigl(I+M\bigr)\quad\bigl(M^{2}=-1\bigr)
\end{equation}
and
\begin{eqnarray}
M_{i}M_{i\pm1}&=&-M_{i\pm1}M_{i},\quad M^{2}=-I,\\
M_{i}M_{j}&=&M_{j}M_{i,}\quad \big|i-j\big|\geq2
\end{eqnarray}
which forms ``extra special 2-group''. Obviously, M is extension of i for $\textrm{i}^{2}=-1$.

An interesting observation is \cite{ge2014yang}:
\begin{equation}
M=-\textrm{i}\hat{C}
\end{equation}
where $\hat{C}$ is the charge conjugate operator in Majorana spinor.
The eigenstates of $\hat{C}$ take the forms
\begin{equation}
\hat{C}\bigl|\xi_{\pm}\bigr\rangle=\mp\bigl|\xi_{\pm}\bigr\rangle,\qquad\hat{C}\bigl|\eta_{\pm}\bigr\rangle=\mp\bigl|\eta_{\pm}\bigr\rangle,
\end{equation}
where
\begin{eqnarray}
&&\bigl|\xi_{\pm}\bigr\rangle=\frac{1}{\sqrt{2}}\bigl(\bigl|\bigl\uparrow\bigr\uparrow\bigr\rangle\pm \textrm{i}\bigl|\bigl\downarrow\bigr\downarrow\bigr\rangle\bigr)\label{EigenKsi},\\ 
&&\bigl|\eta_{\pm}\bigr\rangle=\frac{1}{\sqrt{2}}\bigl(\bigl|\bigl\uparrow\bigr\downarrow\bigr\rangle\pm \textrm{i}\bigl|\bigl\downarrow\bigr\uparrow\bigr\rangle\bigr)\label{EigenEta}.
\end{eqnarray}

Here we would like to give an intuitive interpretation of the operator $\Gamma$ in Ref. \cite{lee2013algebra} by taking a new set of $D_i \,(i=1,2,3)$ in stead of $\gamma_i$, and show how it gives rise to the Majorona
doubling with explicit realization.

We follow the concrete realization for $\gamma_j$ given in Ref. \cite{lee2013algebra}, (in this paper $I$ is $2\times2$ identity matrix)
\begin{eqnarray}
&\gamma_{1}=\sigma_{1}\otimes I,\, \gamma_{2}=\sigma_{3}\otimes I,\, \gamma_{3}=\sigma_{2}\otimes\sigma_{1},&\\
&P=\sigma_{2}\otimes\sigma_{3},&\\
&\Gamma=-\textrm{i}\gamma_{1}\gamma_{2}\gamma_{3}=-I\otimes\sigma_{1}.&
\end{eqnarray}

In our notation, $\gamma_{3}=-\hat{C}$, i.e. (\ref{EigenKsi}) and (\ref{EigenEta}) are eigenstates
of $\gamma_3$. It is easy to find
\begin{align}
\gamma_{1}\bigl|\xi_{\pm}\bigr\rangle&=\pm \textrm{i}\bigl|\eta_{\mp}\bigr\rangle, \quad \gamma_{1}\bigl|\eta_{\pm}\bigr\rangle=\pm \textrm{i}\bigl|\xi_{\mp}\bigr\rangle \label{b_1operation};\\ 
\gamma_{2}\bigl|\xi_{\pm}\bigr\rangle&=\bigl|\xi_{\mp}\bigr\rangle, \,  \qquad \gamma_{2}\bigl|\eta_{\pm}\bigr\rangle=\bigl|\eta_{\mp}\bigr\rangle;\\
\gamma_{3}\bigl|\xi_{\pm}\bigr\rangle&=\pm\bigl|\xi_{\pm}\bigr\rangle, \,\, \quad \gamma_{3}\bigl|\eta_{\pm}\bigr\rangle=\pm\bigl|\eta_{\pm}\bigr\rangle;\\
P\bigl|\xi_{\pm}\bigr\rangle&=\mp\bigl|\eta_{\mp}\bigr\rangle, \, \,\,\quad P\bigl|\eta_{\pm}\bigr\rangle=\pm\bigl|\xi_{\mp}\bigr\rangle;\\
\Gamma|\xi_{\pm}\rangle&=-|\eta_{\pm}\rangle, \qquad\Gamma|\eta_{\pm}\rangle=-|\xi_{\pm}\rangle. \label{Gammaoperation}
\end{align}

In the derivation of (\ref{b_1operation})-(\ref{Gammaoperation}), the relations $\sigma_{1}=(S^{+}+S^{-})$
and $\sigma_{2}=\frac{1}{\textrm{i}}(S^{+}-S^{-})$ have been used where $S^{\pm}=S_{1}\pm \textrm{i}S_{2}.$
To show the importance of $\Gamma$-operator we define new Clifford algebra
$\left\{ D_{i},D_{j}\right\} =2\delta_{ij}$, where $D_{1}=\gamma_{2},\: D_{2}=\Gamma \gamma_{1},\: D_{3}=\gamma_{3}.$
It is interesting to find that

\begin{eqnarray}
D_{j}\bigl|\xi\bigr\rangle&=&\sigma_{j}\bigl|\xi\bigr\rangle,\quad D_{j}\bigl|\eta\bigr\rangle=\sigma_{j}\bigl|\eta\bigr\rangle,\qquad(j=1,2,3)\\
\bigl|\xi\bigr\rangle&=&\Biggl(\begin{array}{c}
\bigl|\xi_{+}\bigr\rangle\\
\bigl|\xi_{-}\bigr\rangle
\end{array}\Biggr),\quad\bigl|\eta\bigr\rangle=\Biggl(\begin{array}{c}
\bigl|\eta_{+}\bigr\rangle\\
\bigl|\eta_{-}\bigr\rangle
\end{array}\Biggr).
\end{eqnarray}

Namely, by acting $D_j$ on$\bigl|\xi\bigr\rangle$ or $\bigl|\eta\bigr\rangle$, the representation is exactly Pauli matrices, i.e. belonging to SU(2) algebra. It can be checked that
\begin{equation}
D_{1}D_{2}=-\textrm{i}\Sigma_{2},\quad D_{2}D_{3}=-\textrm{i}\Sigma_{3},\quad D_{1}D_{3}=-\textrm{i}\Sigma_{1},
\end{equation}
where $\Sigma_{i}$ form the reducible representation of $SU(2)$:
\begin{equation}
\Sigma_1=\sigma_1\otimes\sigma_1, \quad \Sigma_2=\sigma_2\otimes\sigma_1, \quad \Sigma_3=\sigma_3\otimes I.
\end{equation}

The introduced interacting Hamiltonian $H_{B}=-\textrm{i}(\alpha D_{1}D_{2}+\beta D_{2}D_{3}+\kappa D_{3}D_{1})$ can be recast to
\begin{equation}\label{3MFinteraction}
H_{B}=-(\alpha_{1}\Sigma_{1}+\alpha_{2}\Sigma_{2}+\alpha_{3}\Sigma_{3}),
\end{equation}
where $\alpha_{1}=-\kappa,\;\alpha_{2}=\alpha,\;\alpha_{3}=\beta$.
Noting that $D_{1}D_{2}D_{3}=-\textrm{i}I\otimes I$, i.e. trivial. The direct check gives:
\begin{equation}\label{GammaSigmaCommutation}
\Bigl[\Gamma,\Sigma_{j}\Bigr]=0,\qquad(j=1,2,3)
\end{equation}
and
\[
\Bigl[\Sigma_{j},\Sigma_{k}\Bigr]=\textrm{i}\epsilon_{jkl}\Sigma_{l}.
\]
Then the $H_B$ can be written in the form:
\begin{eqnarray}
H_{B}&=&E\overrightarrow{n}\cdot\overrightarrow{\Sigma},\qquad(\overrightarrow{\Sigma}^{2}=I),\label{3MFinteraction1}\\
\overrightarrow{n}&=&(\sin\zeta \cos\varphi,\, \sin\zeta \sin\varphi,\, \cos\zeta),\\
\cos\zeta&=&-\alpha_{3}/E,\quad \tan\varphi=\alpha_{2}/\alpha_{1}.
\end{eqnarray}
Obviously, $\overrightarrow{\Sigma}$ is reducible 4-d representation of SU(2). Explicitly,
\begin{eqnarray}
\overrightarrow{n}\cdot\overrightarrow{\Sigma}&=&M_{1}+M_{2}\nonumber\\
&=&\left[\begin{array}{cccc}
\cos\zeta & 0 & 0 & \sin\zeta e^{-i\varphi}\\
0 & \cos\zeta & \sin\zeta e^{-i\varphi} & 0\\
0 & \sin\zeta e^{i\varphi} & -\cos\zeta & 0\\
\sin\zeta e^{i\varphi} & 0 & 0 & -\cos\zeta
\end{array}\right],
\end{eqnarray}
where
\begin{eqnarray}
M_{1}=\left[\begin{array}{cccc}
\cos\zeta & 0 & 0 & \sin\zeta e^{-i\varphi}\\
0 & 0 & 0 & 0\\
0 & 0 & 0 & 0\\
\sin\zeta e^{i\varphi} & 0 & 0 & -\cos\zeta
\end{array}\right],\\
 M_{2}=\left[\begin{array}{cccc}
0 & 0 & 0 & 0\\
0 & \cos\zeta & \sin\zeta e^{-i\varphi} & 0\\
0 & \sin\zeta e^{i\varphi} & -\cos\zeta & 0\\
0 & 0 & 0 & 0
\end{array}\right].
\end{eqnarray}
Rewriting $M_1$ and $M_2$ in the form of Pauli matrices, we have
\begin{eqnarray}
M_{1}=\cos\zeta\frac{\sigma_{3}\otimes I+I\otimes\sigma_{3}}{2}+\sin\zeta(e^{-i\varphi}\sigma^{+}\otimes\sigma^{+}+e^{i\varphi}\sigma^{-}\otimes\sigma^{-}),\\
M_{2}=\cos\zeta\frac{\sigma_{3}\otimes I-I\otimes\sigma_{3}}{2}+\sin\zeta(e^{-i\varphi}\sigma^{+}\otimes\sigma^{-}+e^{i\varphi}\sigma^{-}\otimes\sigma^{+}).
\end{eqnarray}

Now the meaning of $H_{B}$ is manifest: 4-dimension is quite different from 2-dimension. The "edge block" leads to $M_1$ with superconducting type of Hamiltonian whereas "interior block" $M_2$ is connected with the usual spin chain. It is easy to find the eigenstates of $M_1$ and $M_2$:

\begin{equation}
M_{1}\Bigl|\psi_{1}\Bigr\rangle=\Bigl|\psi_{1}\Bigr\rangle,\qquad M_{2}\Bigl|\psi_{2}\Bigr\rangle=\Bigl|\psi_{2}\Bigr\rangle,
\end{equation}
where
\begin{equation}\label{REigenstate}
\Bigl|\psi_{1}\Bigr\rangle=\left[\begin{array}{c}
\cos\frac{\zeta}{2}\\
0\\
0\\
\sin\frac{\zeta}{2}\, e^{i\varphi}
\end{array}\right],\qquad\Bigl|\psi_{2}\Bigr\rangle=\left[\begin{array}{c}
0\\
\cos\frac{\zeta}{2}\\
\sin\frac{\zeta}{2}\, e^{i\varphi}\\
0
\end{array}\right].
\end{equation}

Acting $\Gamma$ on (\ref{REigenstate}) it yields
\begin{equation}
\Gamma\Bigl|\psi_{1}\Bigr\rangle=-\Bigl|\psi_{2}\Bigr\rangle,\qquad\Gamma\Bigl|\psi_{2}\Bigr\rangle=-\Bigl|\psi_{1}\Bigr\rangle.
\end{equation}
So $\Gamma$ transforms between $\bigl|\psi_{1}\bigr\rangle$ and $\bigl|\psi_{2}\bigr\rangle$
that holds for the same energy. It never occurs in 2 dimensions. Meanwhile,  equation (\ref{GammaSigmaCommutation}) shows that $\Gamma$ commutes with the Hamiltonian $H_B$, which means that $\Gamma$-transformation does not change the property of Hamiltonian $H_B$.
This example shows that operator $\Gamma$ is crucial in leading to Majorana doubling in dimensions $\geq4$. With the new defination of $D_2$,
we should define a new parity operator:
\begin{equation}
P_{B}=\sigma_{3}\otimes\sigma_{2}.
\end{equation}
Direct check gives the complete set of algebra
\begin{align}
&\left\{ D_{i},D_{j}\right\} =0,\\
&\Gamma^{2}=I, \, \left[\Gamma,D_{j}\right]=0, \, \left[\Gamma,H_{B}\right]=0,\\
\label{PBcommuta}
&P_{B}^{2}=I, \, \left[P_{B},D_{j}\right]=0,\\
&\{\Gamma,P_{B}\}=0, \, [P_{B},H_{B}]=0,\\
&\left[\Gamma,\Sigma_{j}\right]=0, \, \left[\Sigma_{j},\Sigma_{k}\right]=\textrm{i}\epsilon_{jkl}\Sigma_{l}, \, \left(j,k,l=1,2,3\right).
\end{align}

It is noteworthy that the introduced $P_B$ in equation (\ref{PBcommuta}) commutes with $D_j$  instead of the anti commuting relation between $P$ and $\gamma_j$. And $P_B$ still anticommutes with $\Gamma$. Acting $P_{B}$ on the eigenstates $\left|\psi_{1}\right\rangle $ and $\left|\psi_{2}\right\rangle$, it follows
\begin{equation}
P_{B}\left|\psi_{1}\right\rangle =\textrm{i}\left|\psi_{2}\right\rangle ,\quad P_{B}\left|\psi_{2}\right\rangle =-\textrm{i}\left|\psi_{1}\right\rangle.
\end{equation}

In such a concrete realization $\Gamma$ plays the essential role. The Hamiltonian
(\ref{3MFinteraction1}) formed by (\ref{3MFinteraction}) looks a typical nuclear resonant model in 4 dimensions. Only the higher dimensions allow the operator $\Gamma$ leading to the doubling
degeneracy.

\bigskip

\noindent \textbf{Majorana doubling in 3-body Hamiltonian based on YBE.}
Now we discuss the interaction of 3 Majorana fermions based on YBE.

It is well known that $\breve{R}_i(\theta)$ describes the 2-body interaction. And  the physical meaning of Yang-Baxter equation is that the interaction of the three bodies can be decomposed into three 2-body interactions:
\begin{eqnarray}
 \breve{R}_{123}(\theta_1,\theta_2,\theta_3)
&=&\breve{R}_{12}(\theta_1) \breve{R}_{23}(\theta_2) \breve{R}_{12}(\theta_3)\nonumber\\
&=&\breve{R}_{23}(\theta_3) \breve{R}_{12}(\theta_2) \breve{R}_{23}(\theta_1).\nonumber
 \end{eqnarray}
Because of the constraint in equation (\ref{YBEangularrelation}), $\breve{R}_{123}$ depends only on two free parameters and has the following form \cite{yu2014factorized}:
\begin{equation}\label{3MFExp}
  \breve{R}_{123}(\eta,\beta)=e^{\eta \left(\vec{n}\cdot\vec{\Lambda}\right)},
\end{equation}
where
\begin{eqnarray*}
 \cos\eta &=& \cos\theta_2 \cos\left(\theta_1+\theta_3\right),\\
 \sin\eta &=& \sin\theta_2\sqrt{1+\cos^2(\theta_1-\theta_3)},\\
 \vec{n} &=& \left(
  \begin{array}{ccc}
  \tfrac{1}{\sqrt{2}}\cos\beta,& \tfrac{1}{\sqrt{2}}\cos\beta,& \sin\beta
  \end{array}\right),\\
  \vec{\Lambda} &=& \left(
  \begin{array}{ccc}
  \gamma_1\gamma_2 ,& \gamma_2\gamma_3 ,& \gamma_1\gamma_3
  \end{array}\right),\\
  \cos\beta &=& \tfrac{\sqrt{2}\cos\left(\theta_1-\theta_3\right)}{\sqrt{1+\cos^2\left(\theta_1-\theta_3\right)}},\\
  \sin\beta &=& \tfrac{-\sin\left(\theta_1-\theta_3\right)}{\sqrt{1+\cos^2\left(\theta_1-\theta_3\right)}}.
 \end{eqnarray*}
Here the parameters $\theta_1$ and $\theta_3$ are replaced by $\eta$ and $\beta$. $\breve{R}_{123}(\eta,\beta)$ is also a unitary operator and describes the  interaction of three Majorana operators.

 We suppose that the parameter $\eta$ is time-dependent and $\beta$ is time-independent in $\breve{R}_{123}(\eta,\beta)$, then the desired 3-body Hamiltonian can be obtained from equation (\ref{SchrodingerEquation}):
\begin{eqnarray}\label{3bodyHamilton}
\hat{H}_{123}(t)&=&\textrm{i}\hbar\tfrac{\partial\breve{R}_{123}}{\partial t}\breve{R}_{123}^{-1}\nonumber\\
              &=&\textrm{i}\hbar\dot{\eta}\left[\tfrac{1}{\sqrt{2}}\cos\beta(\gamma_1\gamma_2+\gamma_2\gamma_3) +\sin\beta \gamma_1\gamma_3\right].
\end{eqnarray}
The constructed Hamiltonian, which has been mentioned in Ref. \cite{alicea2011non,lee2013algebra}, describes the 2-body interactions among the three Majorana operators. It describes the effective interaction in a $T$-junction formed by three quantum wires. In Ref. \cite{lee2013algebra}, it has been shown that the above Hamiltonian, which commutes with emergent Majorana operator $\Gamma=-\textrm{i}\gamma_1\gamma_2\gamma_3$, holds Majorana doubling.  From the viewpoint of YBE,  the intrinsic commutation relation is between  $\Gamma$ and the solution of YBE $\breve{R}_i(\theta)=e^{\theta\gamma_i\gamma_{i+1}}$. It is shown that:
\begin{equation}
[\Gamma, \breve{R}_{i}(\theta)]=0 ,  \quad (i=1,2).
\end{equation}
Indeed, the above commutation relation indicates that emergent Majorana operator $\Gamma$ is a new symmetry of the solution $\breve{R}_i(\theta)$ of YBE. It is due to the decomposition of 3-body interaction into three 2-body interactions via the approach of YBE that the derived Hamiltonian holds Majorana doubling.

The  extended emergent Majorana mode $\Gamma'$ supporting odd number $N$ of Majorana operators \cite{lee2013algebra} is,
\begin{equation}
\Gamma'\equiv\textrm{i}^{N(N-1)/2}\prod_{j=1}^{N}\gamma_j .
\end{equation}
It is easy to check that:
\begin{equation}
\left[\Gamma', B_i\right]=0, \quad (i=1,2,...N-1),
\end{equation}
where $B_i=e^{\tfrac{\pi}{4}\gamma_i\gamma_{i+1}}$ is the generator of the braid group $B_N$. The commutation relation indicates that $\Gamma'$ plays the role of an invariant in the braid group $B_N$.
 
\bigskip

\noindent \textbf{Generation of 2n-qubit GHZ state via YBE.}
Quantum entanglement plays an important role in quantum information theory and has been discussed in both theoretical \cite{Horodecki2008quantum} and experimental \cite{Yamamoto2003experimental,lu2007experimental} aspects for a long time. There are various ways in describing different types of entanglement.  It is also well known that the relationship between Yang-Baxter equation and 2-qubit entangled state as well as 3-qubit entanglement has been discussed in Ref. \cite{kauffman2002quantum,zhang2005universal,chen2007braiding,yu2014factorized}. Here we construct high dimensional matrix representation of solution to Yang-Baxter equation and discuss how it generates $2N$-qubit GHZ state for odd $N$.
In previous section, we present Clifford algebric relation for different Majorana operators,
 \begin{equation}\label{}
\{\gamma_i, \gamma_j\}=2\delta_{ij}.
\end{equation}
It can be used for constructing solution to YBE:
\begin{equation}
\breve{R}_i(\theta)=\exp(\theta\gamma_i\gamma_{i+1}).
\end{equation}
The representation of $\gamma_i$ in the Majorana form is given by:
\begin{align}
 \gamma_{2j-1}&=e^{i\varphi}a_j^{\dag}+e^{-i\varphi}a_j,\\
 \gamma_{2j}&=i e^{i\varphi}a_j^{\dag}-i e^{-i\varphi}a_j.
 \end{align}
Then by constructing Yang-Baxter chain, we find its similarity to 1D Kitaev model.

Indeed, the 4D-matrix representation is equivalent to the Majorana fermion representation under Jordan-Wigner transformation. In other words, we can express $\gamma_i$ by matrix directly. For three operators $\gamma_1$, $\gamma_2$ and $\gamma_3$ satisfying Clifford algebra, its 4D matrix representation has been presented in Ref. \cite{lee2013algebra}:
\begin{eqnarray*}
&&\gamma_1=\sigma_1\otimes I,\\
&&\gamma_2=\sigma_3\otimes I,\\
&&\gamma_3=\sigma_2\otimes \sigma_1,
\end{eqnarray*}
here $\sigma_i$ are Pauli matrices.

What we are interested in is constructing higher dimensional matrix representation of $\gamma_i$. Taking 8D representation as an example,  $\gamma_{i}$ is:
\begin{eqnarray*}
&&\gamma_1=\sigma_1\otimes I \otimes I,\\
&&\gamma_2=\sigma_3\otimes\sigma_1\otimes I,\\
&&\gamma_3=\sigma_3\otimes\sigma_3\otimes \sigma_1.
\end{eqnarray*}
Then the matrix form of emergent Majorana mode $\Gamma$ \cite{lee2013algebra} is,
\begin{equation}
\Gamma=-\textrm{i}\gamma_1\gamma_2\gamma_3=-\sigma_1\otimes\sigma_2\otimes \sigma_1.
\end{equation}
The Hamiltonian supporting three Majorana operators has been defined in equation (\ref{Eq.2}):
\begin{eqnarray}\label{H123}
H_{\textrm{int}}&=&-\textrm{i}\left(\alpha \gamma_{1}\gamma_{2}+\beta \gamma_{2}\gamma_{3}+\kappa \gamma_{3}\gamma_{1}\right)\\
                &=&-\alpha \sigma_2\otimes\sigma_1\otimes I-\beta I\otimes\sigma_2\otimes \sigma_1+\kappa \sigma_2\otimes\sigma_3\otimes \sigma_1.\nonumber
\end{eqnarray}
Obviously, $\Gamma$ commutes with the Hamiltonian $H_{\textrm{int}}$.

\bigskip

Let us extend $\Gamma$ to N sites $\Gamma_i$, which should also satisfy Clifford algebra $\{\Gamma_i, \Gamma_j\}=2\delta_{ij}$. The $\Gamma_i$ has the following form:
\begin{eqnarray*}
\gamma_{3i-2}&=&(\sigma_3\otimes\sigma_3\otimes \sigma_3)^{\otimes(i-1)}\otimes\sigma_1\otimes I \otimes I\otimes I\cdots,\\
\gamma_{3i-1}&=&(\sigma_3\otimes\sigma_3\otimes \sigma_3)^{\otimes(i-1)}\otimes\sigma_3\otimes\sigma_1\otimes I\otimes I\cdots,\\
\gamma_{3i}&=&(\sigma_3\otimes\sigma_3\otimes \sigma_3)^{\otimes(i-1)}\otimes\sigma_3\otimes\sigma_3\otimes \sigma_1\otimes I\cdots,
\end{eqnarray*}
\begin{eqnarray}
\Gamma_i&=&-\textrm{i}\gamma_{3i-2}\gamma_{3i-1}\gamma_{3i}\nonumber\\
        &=&-(\sigma_3\otimes\sigma_3\otimes \sigma_3)^{\otimes(i-1)}\otimes\sigma_1\otimes\sigma_2\otimes \sigma_1\otimes I\otimes I\cdots.
\end{eqnarray}
Then we have:
\begin{equation}
\Gamma_i\Gamma_{i+1}=-\textrm{i}I^{\otimes3(i-1)}\otimes(\sigma_2\otimes\sigma_1)^{\otimes3}\otimes I\otimes I\cdots.
\end{equation}
It is easy to check that $e^{\theta\Gamma_i\Gamma_{i+1}}$ is the $4^3$-D matrix solution of YBE, we denote it by $\breve{R}_i^3(\theta)$,
\begin{equation}
\breve{R}_i^3(\theta)=\cos\theta I^{\otimes6}-\textrm{i}\sin\theta(\sigma_2\otimes\sigma_1)^{\otimes3}.
\end{equation}
By acting $\breve{R}_i^3(\theta)$ on 6-qubit natural basis, such as $|\uparrow\uparrow\uparrow\uparrow\uparrow\uparrow\rangle$, we have:
\begin{equation}
\breve{R}_i^3(\theta)|\uparrow\rangle^{\otimes6}=\cos\theta|\uparrow\rangle^{\otimes6}-\sin\theta|\downarrow\rangle^{\otimes6}.
\end{equation}
This state represents a type of 6-qubit entangled states. In the case of $\theta=\frac{\pi}{4}$, the generated state is 6-qubit GHZ state, and $\breve{R}_i^3(\theta=\tfrac{\pi}{4})=e^{\tfrac{\pi}{4}\Gamma_i\Gamma_{i+1}}$ can be regarded as one braiding operation of two emergent Majorana operator $\Gamma_i$ and $\Gamma_{i+1}$.

\bigskip

Now we generalize the $4^3$-D matrix solution of YBE to $4^{n}$ with $n$ odd. The extended Majorana operator supporting any odd number $n$ of Majorana operators reads,
\begin{equation}\label{ExtendGamma}
\Gamma^n=\Gamma'\equiv\textrm{i}^{n(n-1)/2}\prod_{j=1}^{n}\gamma_j,
\end{equation}
where the constraint of Clifford algebra $\{\Gamma^n_i, \Gamma^n_j\}=2\delta_{ij}$ leads to the odd number $n$. $\Gamma^n_i$ can be expressed as:
\begin{equation}\label{HighDClifford}
\Gamma^{n}_i=-(\sigma_3)^{\otimes n(i-1)}\otimes(\sigma_1\otimes\sigma_2)^{\otimes\tfrac{n-1}{2}}\otimes \sigma_1\otimes I\otimes I\cdots.
\end{equation}
Then we have
\begin{equation}
\Gamma_i^n\Gamma_{i+1}^n=-(\textrm{i})I^{\otimes n(i-1)}\otimes(\sigma_2\otimes\sigma_1)^{\otimes n}\otimes I\otimes I\cdots.
\end{equation}
The $4^n$-D ($n$ odd) matrix representation of solution to YBE is:
\begin{eqnarray}
\breve{R}_i^n(\theta)&=&e^{\theta\Gamma^n_i\Gamma^n_{i+1}}\nonumber\\
&=&\cos\theta I^{\otimes2n}-\textrm{i}\sin\theta(\sigma_2\otimes\sigma_1)^{\otimes n} \quad(\textrm{n\; odd}).
\end{eqnarray}
Consequently, we generate the following state by acting $\breve{R}_i^n(\theta)$ on the $2n$($n$ odd)-qubit natural state $|\uparrow\rangle^{\otimes2n}$:
\begin{equation}
\breve{R}_i^n(\theta)|\uparrow\rangle^{\otimes2n}=\cos\theta|\uparrow\rangle^{\otimes2n}-\sin\theta|\downarrow\rangle^{\otimes2n}.
\end{equation}
When $\theta=\frac{\pi}{4}$, the generated state turns into $2n$-qubit GHZ state for odd $n$.

\medskip

\section*{DISCUSSION}
\label{sec5}

In this paper,  based on the solution of YBE in Majorana form, we discuss the topological phase transition in the derived ``superconducting'' chain and the Majorana doubling in 3-body Hamiltonian as well as the generation of 2n-qubit GHZ-type entangled states.  Unlike the braid operator, the solution $\breve{R}_i(\theta)$ of YBE is  parameter-dependent. Hence the unitary operator $\breve{R}_i(\theta)$ can be used for generating the ``superconducting'' chain and the Majorana doubling in 3-body Hamiltonian. Indeed, the derived chain(\ref{YBEtrivil},\ref{YBEtopo}) describes the braiding transformation of nearest-neighbour Majorana sites for $\theta_1=\tfrac{\pi}{4}$ (or $\theta_2=\tfrac{\pi}{4}$). We also find that the  3-body Hamiltonian $\hat{H}_{123}$ derived from $\breve{R}_{123}$ holds Majorana doubling. From the viewpoint of YBE, the commutation relation  $[\Gamma, \hat{H}_{123}]=0$  can be explained by $[\Gamma, \breve{R}_i(\theta)]=0$ (i=1,2), where $\breve{R}_i(\theta)$ is the solution of YBE. In other words, it is the $\Gamma$-symmetry of $\breve{R}(\theta)$ that leads to the $\Gamma$-symmetry of $\hat{H}_{123}$. The commutation relation can also be  generalized to the extended $\Gamma'$-operator(\ref{ExtendGamma}) for odd $N$ sites, $[\Gamma', B_i]=0$ ($i=1,2,...N-1$), hence $\Gamma'$ is an invariant of the braid group $B_N$.

We present a new realization of Majorana doubling based on emergent Majorana mode and show the role of $\Gamma$ in leading to the doubling degeneracy of $H_B$ intuitively.  We also make use of the extended $\Gamma'$-operator to construct high dimensional matrix representation of solution to YBE. By acting the high dimensional matrix representation of solution of YBE on natural basis, we generate the GHZ-type entangled state.  Thus we conclude that the braiding process of the extended $\Gamma'$-operators corresponds to the generation of GHZ entangled state.  These results may guide us to find much closer relationship between Yang-Baxter equation and quantum information as well as condensed matter physics.

\bibliographystyle{unsrt}

\vspace{3mm}

\indent{\bf Acknowledgments}
This work is in part supported by NSF of China (Grant No. 11475088).

\vspace{3mm}

{\bf Author contributions}
M.L.G. proposed the idea, L.W.Y. performed the calculation and derivation, L.W.Y. and M.L.G. prepared the manuscript, all authors reviewed the manuscript.

\vspace{3mm}

{\bf Additional information}

\textbf{Competing financial interests:} The authors declare no
competing financial interests.

\newpage

\end{document}